# Migration from Copper to Fiber Access Network using Passive Optical Network for Green and Dry Field Areas of Pakistan

## Umar Farooq, Sajid Bashir, Tauseef Tasneem, A.Saboor, A.Rauf

*ABSTRACT—Passive Optical Networks (PON) technology brings an evolution in the industry of Telecommunication for the provisioning of High Speed Internet (HSI) and Triple Play bundled Services that includes Voice, Data, and Video Streaming throughout the world. In Pakistan most of the service providers are offering broadband services on traditional copper OSP (Outside Plant) network since 2000. Demand for the high speed internet and broadband is increasing rapidly, it is desired with great need to migrate from traditional copper based OSP network to PON – FTTx (Fiber To The x) infrastructure. Considering the geographical requirements in Pakistan a scalable fiber network is required which can be optimized as per the user's requirements and demands with high speed bandwidth efficiency, involving the minimum losses and with ideal capital expenditure (CAPEX). In this work a platform for migration from copper to fiber access network with a scalable and optimized PON – FTTx infrastructure in green field and dry field areas of Pakistan have been proposed using Geographic Information system (GIS). In any developing country like Pakistan having the same cultural and geographical topology, this platform can be used to migrate from copper to fiber access network to provide the PON based telecom services. The developed platform for migration from copper to PON based fiber has been studied, planned, and then simulated on a selected geographical area of Pakistan with physical execution that showed better and efficient results with reduction in capital and operational expenditures. A factual plan without ambiguities assists the operators of Pakistan to analyze/forecast bandwidth requirements of an area, optimized network planning along with the in time and efficient deployment.*

## I. INTRODUCTION

In the modern era; demand of the high speed internet (HSI) in the form of Multimedia Broadband (MM&BB) services are growing rapidly in the world [1] that is changing the human lifestyles [2] and the environs of their surroundings due to the smooth and in time services enabling. These rapid technical developments in Broadband services connecting entities, societies, building awareness, generating capital [3], supporting social media and creating opportunities to bring change for betterment of humanity [1], [4]-[5], [6]-[9].Broadband is now regarded as essential to a country's infrastructure, to business and overall competitiveness and is gradually moving closer to being widely recognized as a human right [10].

Key focus of the new era content providers is to digitize the services and to create the rich online experience. There are numerous technologies available in the world with service providers who are competing fast rigorously to provide high speed internet and multimedia broadband services with quality installations.

The evolved technologies for the provisioning of multimedia broadband (MM&BB) services have bounds considering bandwidth, reliability, coverage, and cost [1], [4]-[5], [9],[11].Therefore, the major issue to balance the range of cost, coverage and quality [12] should be formulated for the rapid growth and service provisioning of multimedia broadband technologies epically for an average citizen in developing country like Pakistan as per its cultural and topographic variations. To stay competitive and to satisfy bandwidth demand, telecom operators are permanently improving their network infrastructure with new technologies [6].

The trends of broadband technology in Pakistan are in stark contrast compared to worldwide trends where wire-line leads the broadband market. The pace of broadband growth in wireline is very slow in Pakistan that is very much obvious. The slow growth is due to some factors that need improvement; these factors include [13]:

1) Low (level of) consumer awareness,
2) No coverage of Broadband services,
3) Traffic reduction in broadband services,
4) Low literacy rate,
5) Low local content development,
6) Low computer penetration,
7) Cost of service (Tariff),

Out of the mentioned factors the most important is the coverage area and wire line infrastructure deployment especially for Passive Optical Networks (PON). FiberOptic (FO) is the latest and the most advanced mode of data transmission having the huge bandwidth, interference free, best signal security, fast upgradability, low cost, small size and less weight etc. But at present the PON have a share of only 0.4% in broadband services [9].

Passive Optical Network (PON) technology can be exploited by planning and deployment in a variety of topographic infrastructures which are mainly categorized as FTTx (Fiber to the x) [6]-[7], [12],[14]-[15].

The service providers are facing problems for the maintenance and monitoring of their wire-line networks i.e. copper network, which needs the uplifting, continuous maintenance and rehabilitation to improve the customer satisfaction index [9].

The penetration of the Passive Optical Network (PON) in Access is very slow as compared to the increased bandwidth demand. In Pakistan no any integrated platform is available that can be deployed as per topographic variations and cultural diversity. This platform will offer:

Ease and time saving in Planning, Optimization and Monitoring of PON

1) Low Capital Investment involvement
2) Futuristic

There is a big research work in the network migration from copper to the fiber that has been published [1], [4]-[8], [11]-17], [26-43]. The already researched work needs a lot of tunings and modifications before deployment the same in Pakistan as stated in the problem statement earlier.

It is suggested to build a platform to Migrate from Copper to Fiber Access Network using Passive Optical Network for Green and Dry Field Areas of Pakistan. The platform is proposed because more bandwidth requirement of the subscribers and new customers due to bandwidth hungry application like HDTV, Online gaming, video conferencing etc. also most of the existing copper access network in Pakistan is out aged especially bandwidth demand areas i.e. urban and sub urban areas and bandwidth demand cannot be fulfilled without deployment of PON FTTx access network infrastructure. Major hurdle in copper access network is distance limitation; the signal declines severely when it reaches that limit for specific bit rate that results in increased customer fault ratio [1], [4]-[5], [9], [16]. PON technology is selected being futuristic technology as the access network will be passive and no external power required to any passive component, thus saving also energy resources [1], [16]-[17].

On selected geographic location ArcGIS is used for land base digitization, using the same digitized base map the physical survey will be carried out. Optical-Distribution Network will be planned and designed using ArcGIS upon the validation of data collected during the field surveys. Calculations for power budget and different analysis will be done for Optical-Distribution Network (ODN) optimization aided with simulations. Final optimized Optical-Distribution Network design will be deployed and the results taken by simulated network will be compared with the final results for further improvements if required. Upon the successful execution and deployment the as built diagram will be updated on GIS data for the future use and also for real time monitoring upon integration of GIS system with GOPN network elements.

## II. FIBER BASED ACCESS NETWORK

Total transformation from copper to fiber is not feasible in one go especially in Pakistan. The main reason is the capital investment required for this purpose is very high [1], [4]-[5], [8]. The Fiber-Optic (FO) OSP includes two types of network: 1) Active-Optical Network (AON); 2) Passive-Optical Networks (PON). Fig.1 presents the both PON and AON access networks.

The Active-Optical Networks, the primary copper cables are replaced in steps by introducing active elements in field [1], [4]-[5], [8]. The distribution cabinet in the access local cable network of legacy copper based OSP is replaced with the Optical-Network Units (ONUs). These ONUs are used as intermediate nodes because both Fiber-Optic (FO) cables and copper access network (secondary cables) are terminated on it. These nodes convert the electrical signals in to optical signals and transmit the subscriber traffic to the respective switching Inside Plant (ISP) node using a transmission node. This transmission node is SDH based. ONUs are available in different sizes and comes in both indoor and outdoor configuration [1], [4]-[5], [8].

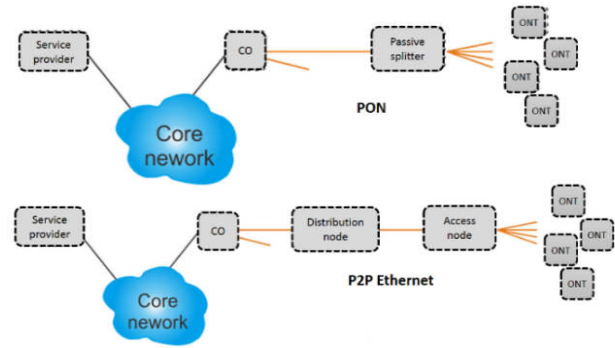

**Fig.1 Active and Passive Optical Network [18]**

The other one is passive optical network (PON) uses point-to-multipoint fiber to the premises in which unpowered optical splitters are used [18]and are widely deploying worldwide. The name passive depicts that there is no available electrical source in OSP or no any active node is required in the fiber access network for service provisioning to the subscribers.

Currently, the major access network in Pakistan is based upon copper which is required to be transformed into Fiber Optic (FO) access network. Fig.2 showing a typical optic fiber based OSP in which different category customers like corporate, residential, enterprise and building etc. are connected with access network using Fiber Optic (FO) s.

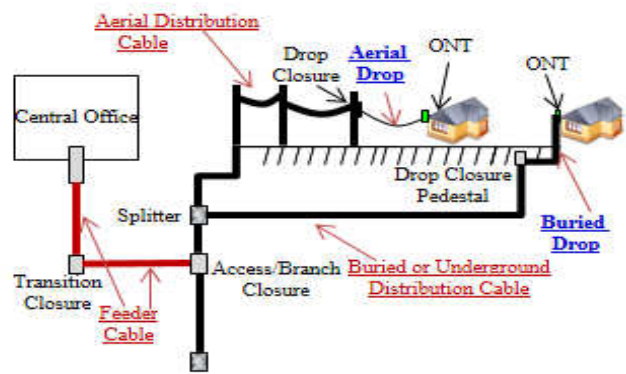

**Fig.2 Fiber based Outside Plant (OSP)**

## III. GIS FOR TELECOMMUNICATION

GIS is an information system that is used to input, store, manipulate, retrieve, analyze and generate geographically-referenced data or geospatial data, take in decision making for planning and management of telecommunication, natural resources, land use, urban services and facilities, transportation, environment and other related administrative records.

### A. GIS Structure

GIS structure is based on layer type, which is integrated all together to get the desired data. Out of these layers Base Map is most critical and the important requirement of the GIS.

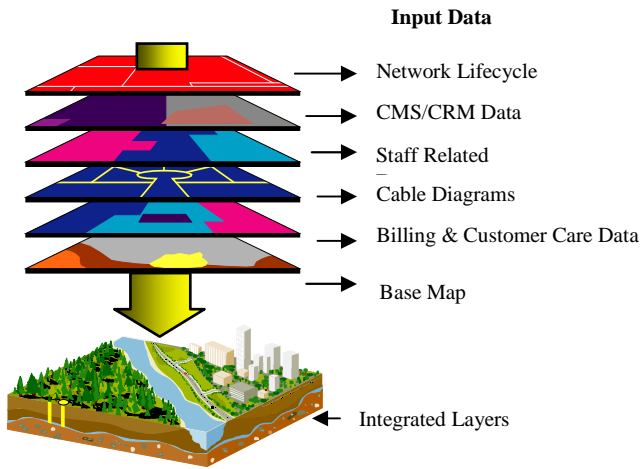

**Fig.3 GIS Structure**

### B. Base Map Preparation

In GIS the creation of base map is the basic requirement to overlay any other data on it. This means before planning and design Fiber Out-side plant using GIS we have to prepare the base map of that specific area. After the creation of base maps all of the fiber related layers will be created and used for access network planning and designing with reference to that base map.

ArcGIS is the well-known GIS application used to create, manipulates, analysis and output creation of data and information. ArcGIS latest version 10.2 is released and available in the market.

### C. Geo Referencing

Before going to start the digitization of Base Map, some basic inputs are mandatory that includes:

- An image of the site map as .tif, or .png or .jpeg file
- Or a rich satellite image
- Map or image is Georeferenced at first step
- Google Earth
- Arc Map 10.0 or latest

Geo-Referencing is the Pre digitization step, it means that we need to make our data (shape file, imagery or map) understand its geographic reference on earth; it's just like playing with puzzles. The steps are:

1) First to collect ground control points or (x, y) co-ordinates and then assign these coordinates tothe site image or map that is being georeferenced.

2) GPS can be used to collect the coordinates by physical survey of the site, or can be picked up by using Google Earth.

3) Picking the coordinates by using Google Earth is very easy, just need to place mark for the corner or building of the area for which you want to collect coordinates as shown in Fig.4.

4) Geo-referencing .CAD (AutoCAD drawing), export the CAD drawing to jpeg format by opening it in AutoCAD or Arc GIS. After this the same process as above is used to collect or assign coordinates for the selected ground control points.

5) Geo-referencing a Map or Imagery, using the same technique as of .CAD drawing; map should be in .tiff or .jpeg format.

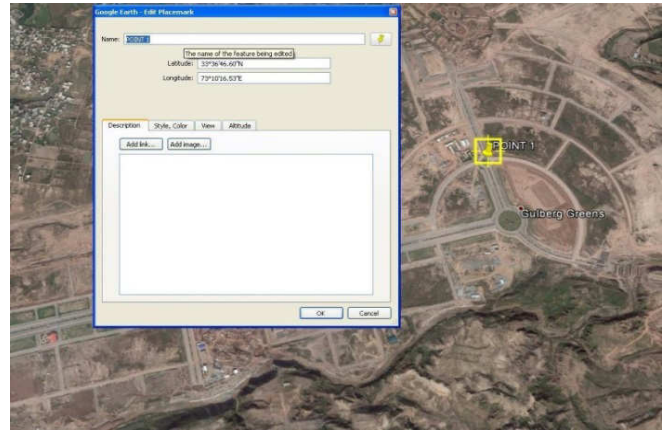

**Fig.4 Taking (x, y) coordinates from Google Earth**

6) Open Arc GIS application, from menu bar click customized tab and select tool bar then select Geo-referencing as shown in Fig.5

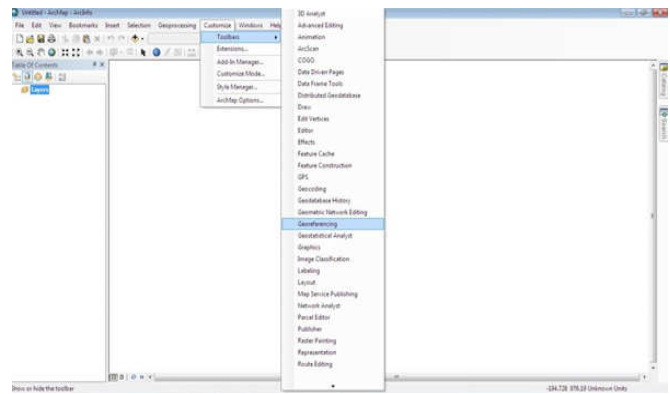

**Fig.5 Geo-referencing in Arc Map, Step (1)**

7) Add the JPEG image or satellite imagery in Arc Map,in the beginning assign that image proper coordinate system that will be available to all data sets of allied Telecom layers and Land base feature class. It is recommended to import the current geographic coordinate system or get the projection from Arc catalog, WGS 1984 43N is used for Pakistan.

8) After this assign each point that is already selected as ground control point to its individual geographic coordinates taken from the Google Earth, for the this purpose first left click on the corner or edge on the image and then right click and click input x and y coordinate value as shown in Fig.6. At minimum four points are needed for effective geo referencing.

9) After assigning coordinates to all of four points click update geo referencing, by doing so image will be automatically rotated and moved to its original coordinated location and is ready to be digitized.

10) Separate layers for the base map are created for separate data set along with allied attribute info which includes layers of zone or sub zone, water bodies, road, parcel (polygon) shown in Fig.7.

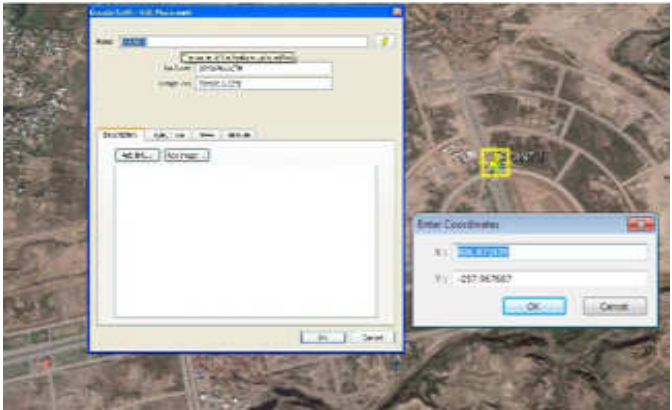

**Fig.6 Geo-referencing in Arc Map, Step (2)**

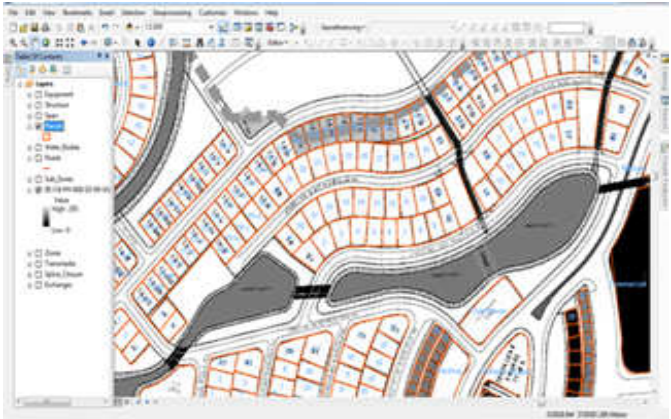

**Fig.7 Digitized Roads, Parcels or polygon**

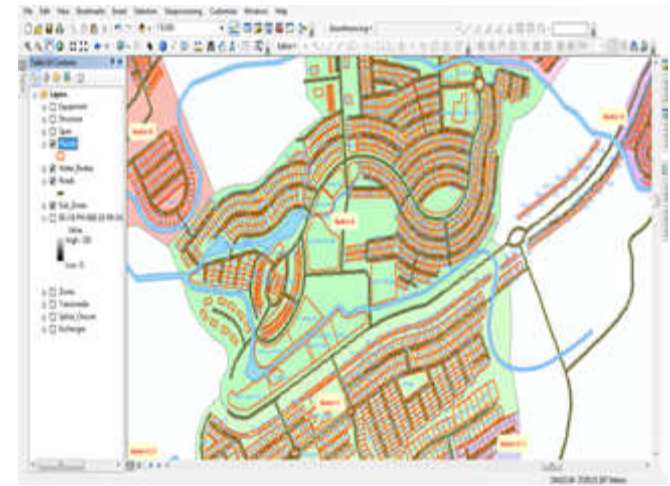

**Fig.8 Base Map**

11) Fig.8 is final and ready product for survey, planning, designing and optimizing, analyzing along with budget calculation for GPON infrastructure. In a similar approach, layers for GPON infrastructure will also be created that includes PVC duct routes for Fiber Optic (FO) OSP, Structure (for hand holes, manholes and JBs), and equipment (for FDH, FAT and OLT etc.), Feeder cables and Distribution cables or any other if required.

## IV. PON ACCESS NETWORK DESIGNING

Fiber to the Home (FTTH) is network having end to end fiber in access network that has become ultimate requirement for any service provider in wire line business. FTTH could be designed and deployed in two topologies:

- Point-To Point (PTP) Network,

- Point-To Multi Point (PTMP) Network.

In PTP a dedicated fiber for each tenant whereas PTMP used the fiber sharing among a tenant group, typically 32 subscribers shown in Fig.9 (a) and (b).

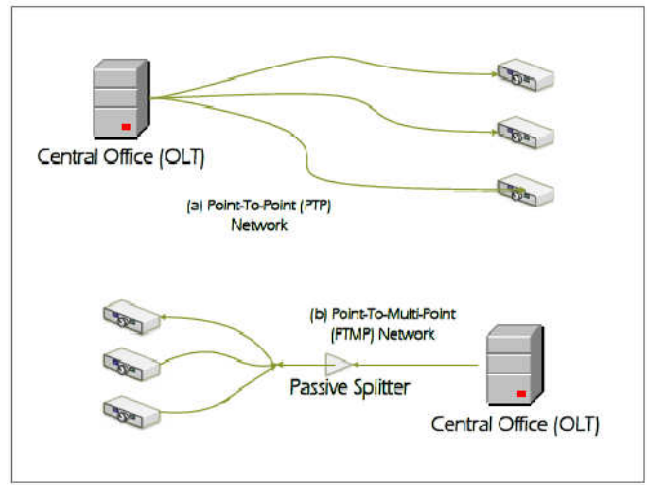

**Fig.9 PTP and PTMP Networks**

In PTP network, there is an individual dedicated fiber with a dedicated port of OLT for a single user. Only one ONT is directly connected to the OLT PON port, requires a huge investment and not a good approach. It is suggested to keep spare capacity in feeder and distribution cables to fulfil such minute demands. However, in PTMP PON network the optical reach is less than that of PTP PON structure that has no splitters installed in it. A PON port could cover a distance of 20 Km from the OLT to cover most of the area. PTP PON networks should be deployed in the remote areas where OLT placement is not economical but demand exists.

In PTMP PON access networks, PON port is shared among the subscribers using the splitters. A single OLT PON port could be shared between 2 to 128 users depending upon the bandwidth requirement and splitters availability. The GPON access network that we are developing for Pakistan will promise asymmetrical communication with 2.5 Gb/s downstream and 1.25 Gb/s upstream. GPON technology supports Ethernet, ATM and WDM by using superset multi-protocol layer. GPON Encapsulation Method (GEM) a transport protocol layer is used to support both Ethernet as well as ATM protocols.

### A. Optical Distribution Network (ODN) Planning

The GPON ODN designed is based upon four basic sections of PON networks that include:

OLT/CO Planning
Splitter Planning
Route Protection/Diversity Planning
FTTx planning

### B. OLT/CO Planning

Location finalization for Optical Line Terminal (OLT) has great importance, Big impact of OLT location on increase or decrease in investment. In telecom network the central office (CO) is Inside Part that houses the OLT and Optical Distribution Frame (ODF). The ODF is installed such that the all feeder cable could be terminated on it. An ODF should not be expanded or any expansion in ODF must be done in same room. An Optical Distribution Network (ODN) is planned for serving multi services. The ODF must provide flexible arrangement to manage fibers (jumper, pigtails, and feeder) with protection for bending radius. Legacy copper cables Main Distribution Frames (MDFs) and copper cable chambers are not required for FiberOptic (FO) cables, since existing standard cable trays and channels can be used to route fiber cables to ODF. OLT should be installed in central office (CO) in standard rack with termination at front side. The OLT rack comprise of 2 to 3 sub racks, each with 16 GPON cards, 4 or 8 PON ports per GPON card.

Physical reach is defined as the maximum physical distance between the ONT and the OLT. In GPON,two options are defined for the physical reach: 10 km and 20 km. It is assumed that 10 km is themaximum distance over which FP-LD can be used in the ONU for high bit rates such as 1.25 Gbit/sor above [19], but it may reduce with the available components. The reach range of GPON could be increased or decreased by changing the location, reducing or adding the splitter levels. Fig.10 defines the Central Office (CO) / OLT placement. OLTs are normally installed in existing exchanges for smooth transformation of copper; this will reduce the cost by re-utilizing the existing resources. Existing cable routes of access and junction including existing civil structures e.g. Manholes and hand holes should be used to extend feeder fiber cables to proposed FDH locations.

OLT should be proposed at place with the concept to feed the locality within its physical reach. Green Field or remote far areas should be feed directly by extended PON port. Boundaries of two OLTs should not be overlapped as it is the capital wastage. The customers within 500 meters should be feed directly from OLT without any FDH or FAT placement. The OLT capacity planning depends upon the potential business demand and expected demand of tenants.

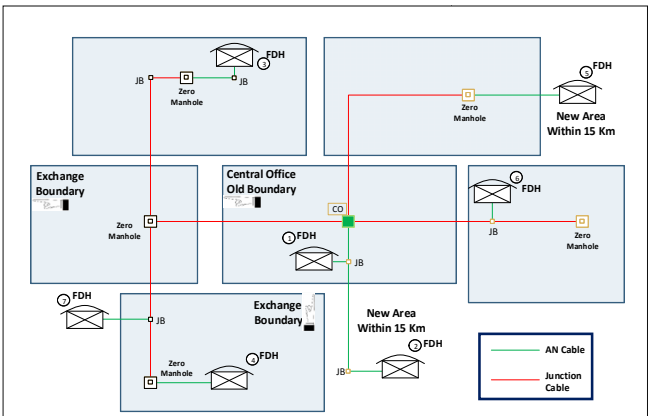

**Fig.10 OLT Planning**

### C. Splitter Planning

Splitter planning depends on the 1) Splitter n-level setting 2) centralized setting 3) Dispersed setting. Fig-11 showing the said three setting, splitter could be used in 2 levels that is 1st splitter of size 1:2 or 2:2 could be installed in central office and the 2nd splitter of size 1:32 or 2:32, 1:16 or 2:16, 1:8 or 2:8, 1:4 0r 2:4 etc. could be installed in FDH or in FAT. Increasing the splitter size will result reduction the limit of bandwidth with increased number of tenants. A PON port supports a maximum of 64 customers delivering the average bandwidth of approx. 38 Mbps per user. However, in case of deployment single level splitting having

1:64 or 2:64 splitters, PON port reach is increased considerably but number of users and bandwidth per users remains same.

Centralized setting of splitters is useful for small coverage area and few splitters are required with minimum of splitter cabinets whereas in decentralized setting, the splitters are installed suing the mounted micro ODFs near to a group of customers, resulting in increased number of splitter cabinets also PON port usage is poor.

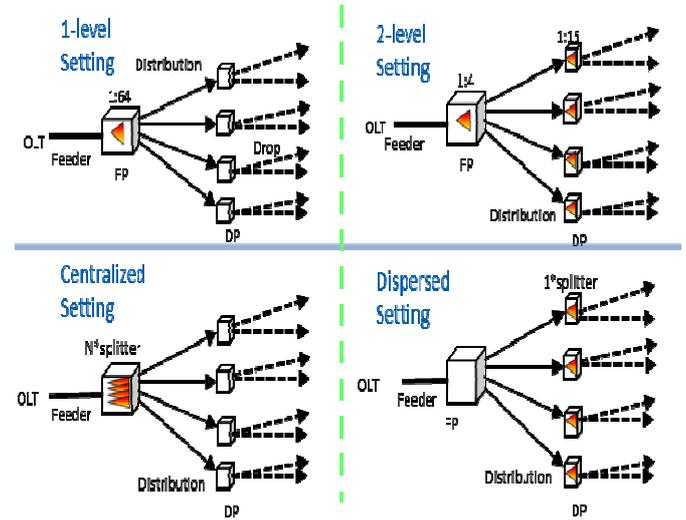

**Fig.11 Splitter Planning**

### D. Telecom Resilience

Network diversity ensures the customer protection for uninterrupted services. In diversity, there is a compromise between cost and uninterrupted service availability. There are three ways to achieve link protection:

• Type A Protection: There is no automatic protection available in both distribution and feeder.

• Type B Protection: Routing protection is provided in Type B, only feeder cables are protected by using of 2: N splitter like 2:2, 2:16, and 2:32 etc., two separate feeder cables either from same OLT card or from different OLT card to the Fiber Distribution Hub (FDH)

• Type C Protection: Maximum protection by drawing both feeder as well as distribution cable in mirrored, it is not an economical solution.

### E. Fiber Flooding-FTTx

Fiber To The-x is actually the fiber flooding from OLT to ONT. Migration from copper access network to fiber access network is a step by step transformation. It is based on the concept of fiber in copper out (FICO) with FTTH is the ultimate requirement. FTTx in different flavors being deployed in the world wide, depending upon the site situation. Fig.12 showing the different variants of FTTx that includes:

• FTTB – Fiber to the Building
• FTTC – Fiber to the Curb
• FTTH – Fiber to the Home
• FTTM – Fiber to the Mobile (fiber backbone to mobile BTS towers)

# Migration from Copper to Fiber Access Network using Passive Optical Network for Green and Dry Field Areas of Pakistan

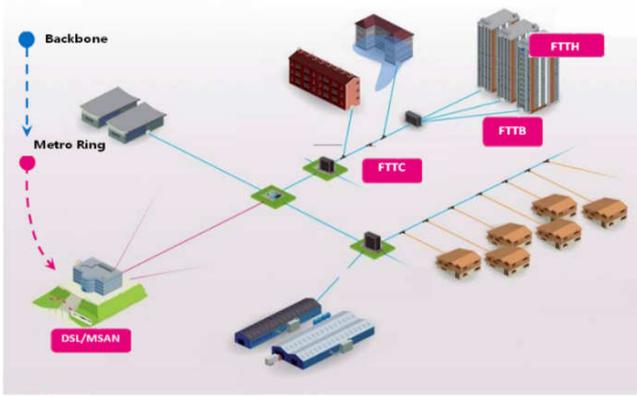

**Fig.12 FTTX Planning – Fiber Flooding**

## V. GREEN FIELD AND DRY FIELD AREAS PLANNING

The fiber count from the central office OLT shall be with spare capacity and of suitable size that ensure future requirements of fiber capacity. In planning or forecasting 25% more fibers rounded to the nearest next cable size is selected e.g. for demand count of 20 fibers, at least 24 fiber cable is planned. While in fiber cable planning process, section lengths of Manhole, Hand holes should be considered keeping in mind the cable drum length to avoid the unwanted joints or splices that can increase the loss and results in poor grade of service. To increase customer base and for business development, the OSP design should be capable enough to meet with the hidden demand, therefore, number of direct fibers should be accompanied while planning for main fiber cables to retain room for the provision of PTP services or dedicated fiber provisioning. The spare capacity that is available in junction or existing PVC cable routes should be utilized to access a remote green field area to connect a FDH or FAT with the OLT.

Requirements of duct routes and related civil structures (hand holes, manholes, joint chambers or boxes etc.) significantly reduced in the PON OSP network due to the fiber characteristic like high bandwidth carrying capacity and small in size and most importantly due to fiber splitting. It should be ensured that the main feeder fiber cables should not be accessed frequently to divert or put through fibers. The drop fiber closures and splicing trays should be placed inside joint boxes located near to group of villas or as per the field requirement, again cable drum lengths must be considered while planning for distribution cables.

The OSP deployment in the dry field area is not a simple process; it is difficult as compared with development in the green field areas. Therefore migration from copper to FTTH in the dry field areas should be done in phases and on case to case basis in accordance to resource availability. Deployment of FTTH in the urban areas should be more preferred over the remote areas. Existing laid fiber cables, cabinets and the different telecom civil structures of the network must be considered during the planning and deployment process of GPON network to reduce the cost.

### A. Splitter Calculation and Cable Sizing

In PON access networks a dedicated fiber is required by every splitter and a dedicated cable from splitter to ONT. Fiber network have a typical life of 20 years, therefore planning is done to provide a network which could support the potential services and future demands. The split ratio and localization of splitters depend on the engineering plan and the mapping distribution of the premises [20]. Further, fiber cost is less than the cost of PVC duct space, therefore it is suggested that to provide the number of fibers to cater for 20 years tenants requirements. Splitters per cabinet are calculated by total number of tenants dividing by the split ratio.

Total number of Splitter = No. of Tenants / split ratio

The calculated number of splitters can be used to decide the size of optical fiber cable, maximum possible expansion and outstanding fiber.

### B. Distribution and Drop Fiber Cables

The distribution cables with loose tube having different sizes 96F, 48F, 24F, 16F or 8F should be used or any combination of different sizes depending on the villas locations, numbers and grouping.

The drop fibers mostly used in the size of 2F core. Enclosures should be capable to accommodate 8 to 24 drop cables. In the small building areas or villas the FDH is located as outdoor and distribution cables should be considered as the outgoing cables. It is recommended to splice one fiber of drop cable with the distribution cable to extend the services to the ONT while keeping the second fiber spare (stumped). The stumped fiber could be used for the maintenance purpose or for an additional connection to a portion of same villas or building.

The scenario of overhead or aerial distribution cables, fiber cables are erected in similar fashion, with distribution fiber cables from FDH and the drop fibers from the enclosures. The fiber cables in access network must be properly labeled and arranged. The extended drop fiber cables in micro ODF installed for single villas in dry and green fields. All the apartments and flats must be pre cabled up to the installed micro ODF ensuring that there will be no splice in between the tenants and micro ODF. Fig.13 showing the fiber cable distribution from FDH to each user.

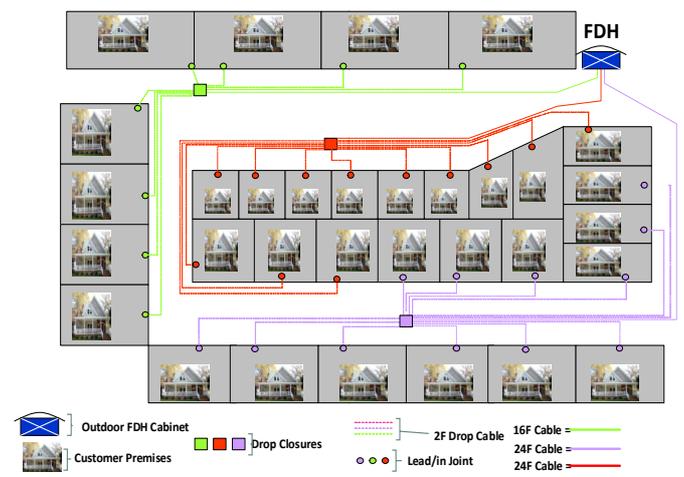

**Fig-13 Cable Distributing from Cabinet**

### C. Developed PON Infrastructure

Fig.14 (a) to (d) describes the ODN designing for different scenarios extend PON based services to different category of users in both green and dry field sites starting from the OLT in central office till customer's premises. The ODN design is finalized by considering all the planning requirements including central office, splitter, route protection/diversity and FTTx – fiber flooding planning.

**Scenarios – I: ODN Design for Villas with Outdoor FDH**

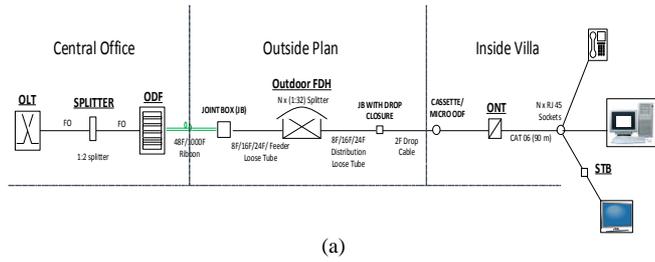

(a)

**Scenarios – II (a): ODN Design for High Rise Building with Indoor FDH**

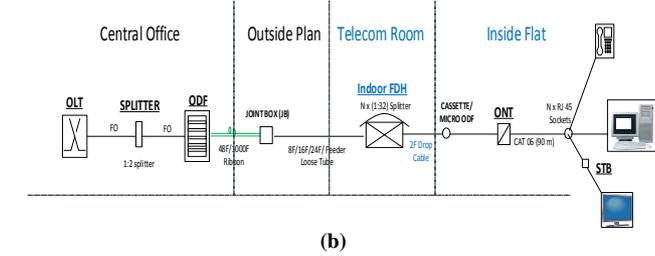

(b)

**Scenarios – II (b): ODN Design for Small Building (up to 5 Floors) with less than 32 subscribers with Indoor wall mounted FDH**

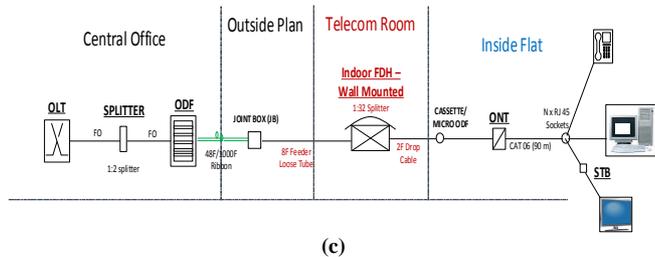

(c)

**Scenarios – II (c): ODN Design for Small Building (up to 5 Floors) with less than 32 subscribers with Outdoor FDH**

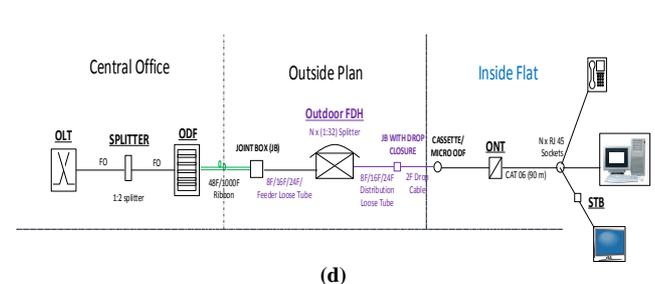

(d)

**Fig.14 (a) – (d) Optical Distribution Network – (ODN) Design**

### D. Digitization of ODN Using GIS

The ODN access network planning and designing using the GIS is the utmost requirement. Since, GIS gives help to analyze the area, to calculate the existing potential demand as well as the forecasted and anticipated demand. GIS helps to speed up the designing and planning process by boosting the survey activities. Separate layer for each OSP network component should be prepared and are integrated together with base map to get the final product. Fig.15 to 19 explaining the digitization of ODN access network planning using GIS.

- First step to have digitized the base map and to draw the sector boundaries. The exact location for OLT and FDH

along with their service area must be defined by forecasting the ultimate users shown in Fig-16.

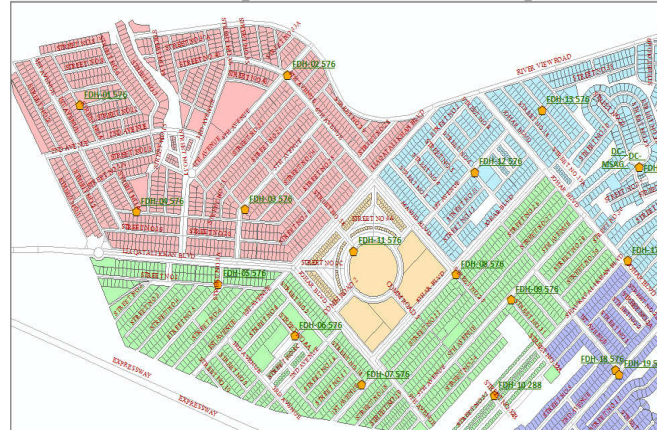

**Fig.15 FDH proposed locations**

- Duct planning for feeder cables connectivity from central office OLT to FDH is done according to the FDH boundaries by considering the cost impact. The placement of allied civil structures Manholes, Hand hole and Joint Boxes for PVC duct should also be planned along, shown in Fig.16.
- Once PVC duct planning for feeder cables is completed, the feeder fiber cables are calculated as per the splitter requirements.

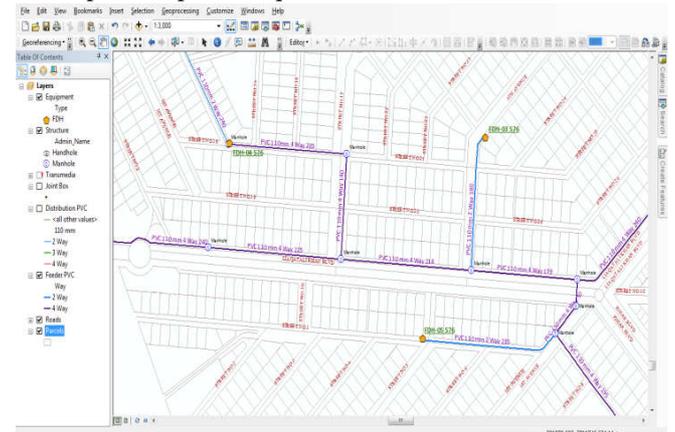

**Fig.16 Duct and civil structure planning**

- Further identify the locations for joint box to connect each user with distribution cable via drop fiber cables; it is time taking and most critical activity shown in Fig.17.

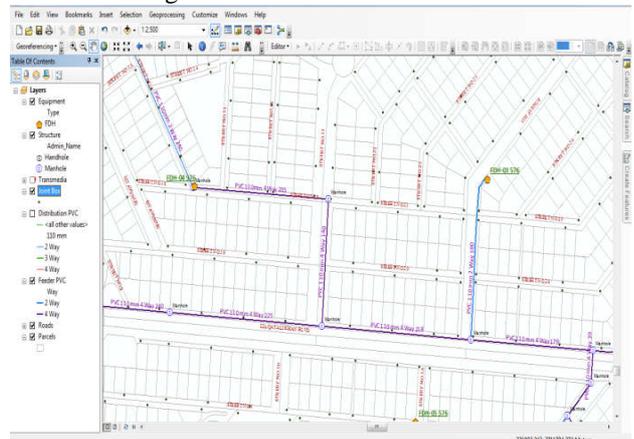

**Fig-17 Proposed location of joint boxes and enclosures (small dots are joint boxes)**

# Migration from Copper to Fiber Access Network using Passive Optical Network for Green and Dry Field Areas of Pakistan

- When the joint box locations have been finalized, distribution PVC ducts and fiber cables could be planned as shown in the Fig.18.

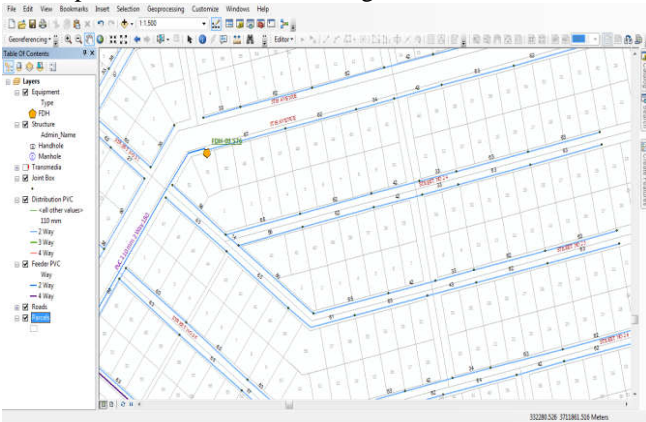

**Fig.18 Distribution network planning**

- At final, all the layers (FDH/FAT locations, its feeding areas, Feeder PVC ducts and fiber cables, distribution ducts (if required) with distribution fiber cables and joint boxes location) are once planned, all the layers could be integrated to get the final product. All the network deployment with development should be done using this planned GIS based ODN design as shown in Fig.19.

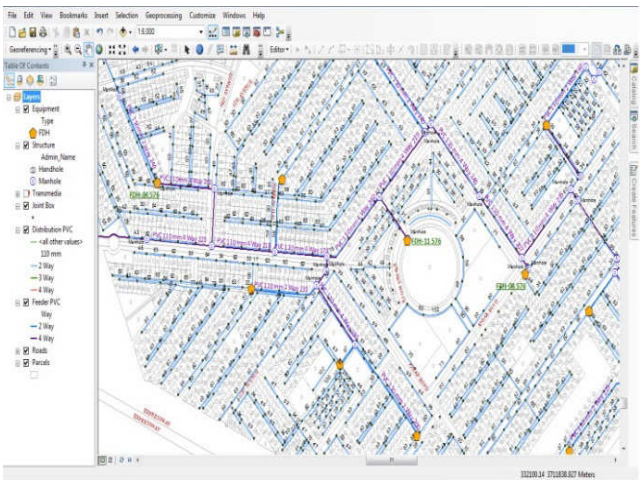

**Fig.19 Integrated GIS layers (ODN Design)**

### E. Power-Budget Calculations of ODN

The ODN design is incomplete and not ready for deployment until and unless end to end loss calculation is done and provided with the ODN design. ODN planning with GIS helps us to complete an optimized design but still required power budget analysis. Power budgeting is the end to end calculation of total maximum loss that an ODN design can offers to each user. In PON system the optical signals are transmitted by the transceiver with a minimum and a maximum possible loss threshold, service could not be provisioned if the single loss level will remain in between the defined threshold values.

GPON optical transmission is based on class B+ standard, according to ITU-T G.984. The optical threshold for the total loss should remain between 13dB and 28dB [21]. ONT will not receive the service if the received signal strength is not lie in between the specified range. Attenuator will be used if the received loss is less the 13db (<13db).

The power budget calculations must be done at the designing stage that gives the analysis of maximum reach ability of an OLT PON port. On the later stage, power budget is calculated at the time of service provisioning to ensure standard grade of service for each user. However, at planning stage, it is not possible to calculate power budget for each and every possible user, with experience, it is learnt that the power budget loss calculations must be done for the farthest tenant, as if the service is perfectly extendable to a distant user then obviously it could be made available to rest of the users.

The calculations for power budget are done by summing the all losses offered by each and every network component. Following formula has been derived for power budget loss calculation upon the finalization of ODN:

| Total Loss of ODN | = | **Sum of all component loss (L)** |
|---|---|---|
| $\sum (L)$ | = | *Addition of Summation of all losses offered by fiber cable (transmission Loss), offered by physical connectors (connection), offered by splatters (splitting loss), offered by splices (splicing loss),and Engineering margin* |
| $\sum (L)$ | = | $\sum TL + \sum CL + \sum SL + \sum SPL + EM$ **(1)** |
| $\sum TL$ (per km) | = | *Transmission Loss cables is offered by the optical fiber cables and measured as per kilometer (km), generally calculated for the smallest wavelength as it gives the maximum loss.* |
| $\sum CL$ (each) | = | *Loss offered by each connector is calculated.* |
| $\sum SL$ (each) | = | *Loss offered by the splitters, different splitters have different insertion loss, larger the spit ratio bigger the loss.* |
| $\sum SPL$ (each) | = | *Loss of each splice (fusion or mechanical)* |
| EM (each) | = | *3 dB engineering margin is kept for and for operation and maintenance (O&M) and future expansion purposes.* |

## VI. SIMULATIONS AND RESULTS

The Bill of Material (BOM) is required for the calculations of power budget, so that comparison could be carried out between the theoretical and practical values acquired upon physical network deployment of planned platform in selected area of Pakistan.

The theoretical insertion losses for different ODN components are commonly known and shown in Table- I

Table- I

| Description | Theoretical Values (dB) | Practical Values (dB) |
|---|---|---|
| Transmission Loss | 0.35 | 0.36 |
| Splicing Loss | 0.1 | 0.05 |
| Connector Loss | 0.2 | 0.21 |
| 1:64 splitter loss | 19.7 | 19.8 |
| 1:32 splitter loss | 17 | 17.13 |
| 1:16 splitter loss | 13.5 | 13.72 |
| 1:8 splitter loss | 10.5 | 10.69 |
| 1:4 splitter loss | 7.2 | 7.15 |
| 1:2 splitter loss | 3.5 | 3.48 |
| Engineering Margin | 3 | 3 |

**Table. I Comparison of Theoretical and Practical Loss values**

Practical scenarios are devised to simulate the developed platform by using theoretical and practical values. The practical scenarios are devised with of modeled structures mentioned in section 5-C of this document.

The simulation is translated in to simple mathematical model and calculations are done using total loss equation derived in section V-E, all the mathematical calculations are done using Microsoft Excel.

Assumptions (All Scenarios): Capacity of each PON port is to serve maximum of 64 users, as the OLT model with maximum split ratio of 64. A 3dB loss as engineering margin is included for O&M purpose.

**Scenarios – I secure average bandwidth of 37.5 Mb/s per Tenant using 1-level of splitting.**

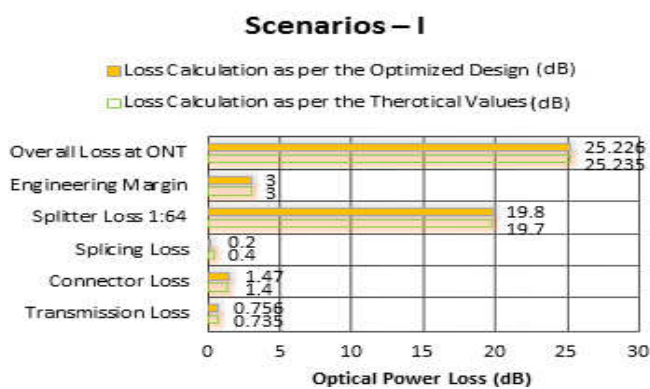

**Fig.20 Results related to scenarios – I**

Result: The proposed platform with this scenarios showing reduction in total loss by 0.036%. 37.5 Mbps maximum bandwidth could be easily made available at a distance of 9.8 km. Using single level or 1-level splitting, there is increase in the reach ability but it is only effective for multi-story or high rise buildings.

**Scenarios – II securing average bandwidth of 37.5 Mb/s per Tenant with 2-level of splitting**

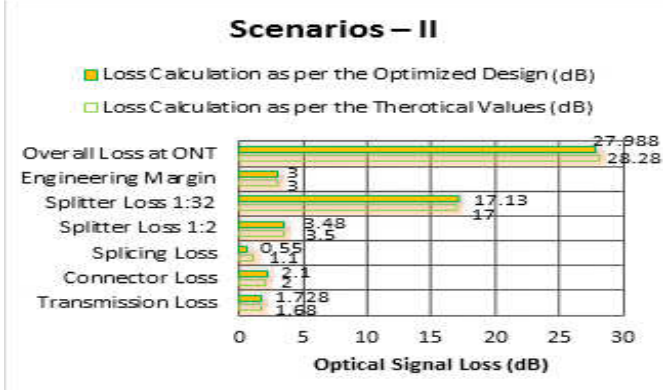

**Fig.21 Results related to scenarios – II**

Result: The proposed platform using two levels splitting shown reduction in overall loss by 1.033%. The maximum bandwidth of 37.5 Mbps could be easily made available at a distance of 4.8 km.

**Scenarios – III securing average bandwidth of 37.5 Mb/s per Tenant with 3-level of splitting**

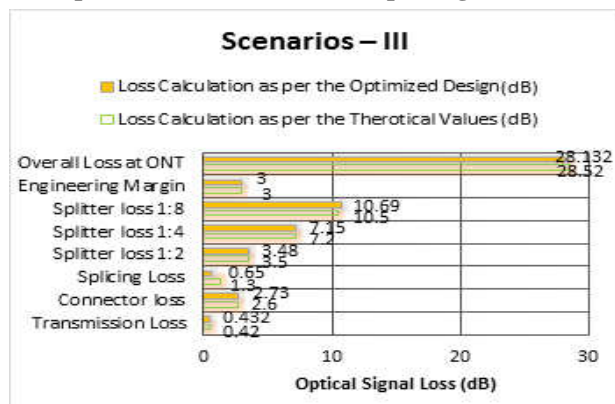

**Fig.22 Results related to scenarios – III**

Result: The proposed platform of three levels splitting the reduction in overall loss by 1.36 %. The maximum bandwidth of 37.5 Mbps could be easily made available at a distance of only 0.8 km. The loss is improved, but due to less reach ability, 3-level of splitting is not recommended.

**Scenarios – IV Securing average bandwidth of 75 Mb/s per Tenant with 2-level of splitting**

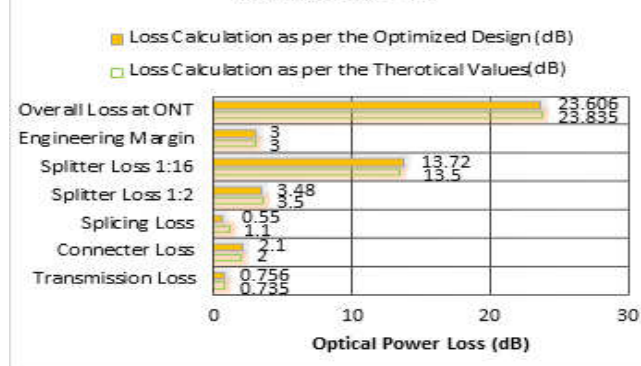

**Fig.23 Results related to scenarios – IV**

Result: The proposed platform in this scenarios shown reduction in overall loss by 0.961%. The average maximum bandwidth of 75 Mbps could be easily made available at a distance of 14 km, however half of the PON port capacity is compromised.

## VII. CONCLUSIONS

1) Section V-D (Fig-16 to 20) shows how the Geo Graphical Information (GIS) could be used to design optical fiber based Outside Plant. The optical distribution network designed on GIS will be an efficient and optimized network required by a service provider to fulfil customer expectations.

2) The proposed planning guidelines together with the equation of total loss:

$$\sum(L) = \sum TL + \sum CL + \sum SL + \sum SPL + EM$$

Will help the operator to develop desired platform for the OSP using PON infrastructure for green and dry field areas of Pakistan.

3) Section VI (Fig-21 to 24) shows that developed platform will not only equipped the service providers of Pakistan with an optimized way of planning but it will also help them to perform:

i. Bandwidth analysis and its forecasting, the actual requirement of the customers can be easily identified for futuristic planning by effective assortment and anticipation of forecasted tenancy. Further, it will support the planner's tocalculate the bandwidth requirement of each and every year along with the bandwidth requirement of the system.

ii. Scalable optimized network designing, the optical distribution network based on developed platform can be easily expanded if additional demand is raised in the same area. Also, PON based services could be easily extended by increasing the reachability of the PON ports.

iii. Deployed of OSP with minimum CAPEX by maximizing quality along with efficient and time saving development.

4) The developed platform could extend additional support in:

i. Network re-engineering, the infrastructure based on this platform is future proof and can incorporated any technological advancements that is same optical access network will be used with new Inside plant equipment's.

ii. Fault localization and management, as the network is available in digitized form and if physical changes and amendments are updated in the already digitized network then fault tracing and its rectification is possible.

iii. Marketing of new services, order booking for new customers is possible by confirming resource availability with the help of updated digitized network footprint.

## VIII. FUTURE WORK

Having seen the results given above, it is evident that the world is going towards digitization. However, there is a need to break the silos so that information sharing process could be speed up. Therefore, integration of different platforms is required, it is not necessary that all these discrete platforms are directly linked to each other instead they are inter depend to achieve the same cause.

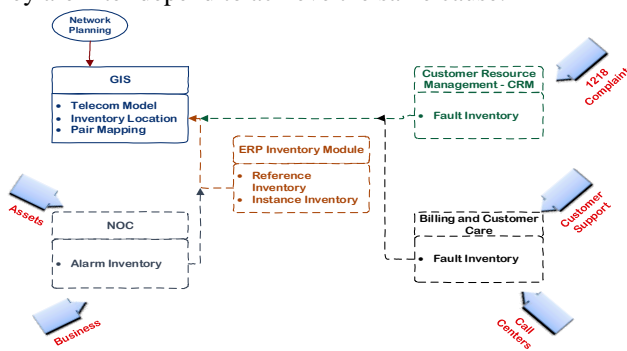

**Fig.24 GIS Integration**

Fig.24 shows an example of indirectly inter depended platforms that are working for single purpose that is to achieve business excellence. The platforms shown are National Operation Center (NOC) which monitors all active network elements to ensure normally service delivery to customers, Billing and Customer Care (Bn CC) captures customer demands and divert it to concerned departments for service provisioning, Customer Resource Management (CRM) register customer complaint and refers it to concerned team for rectification, Enterprise Resource Planning (ERP) an inventory management system, and Geo graphic Information System (GIS).

In the coming future, it will be a requirement that all of such platforms could be integrated altogether with the help of GIS system. The integration will results in the real time monitoring of customer complaints, new customer requests, and NOC alarms with geo-graphical information.